\documentclass[aps,preprint]{revtex4}%
\usepackage{amsfonts}
\usepackage{amsmath}
\usepackage{amssymb}
\usepackage{graphicx}%
\providecommand{\U}[1]{\protect\rule{.1in}{.1in}}
\begin{document}
\title{Genuine Bell locality and its maximal violation in quantum networks}
\author{Li-Yi Hsu}
\affiliation{Department of Physics, Chung Yuan Christian University, Chungli 32081, Taiwan}
\begin{abstract}
In $K$-locality networks, local hidden variables emitted from classical
sources are distributed among limited observers. We explore genuine Bell
locality in classical networks, where, regarding all local hidden variables as
classical objects that can be perfectly cloned and spread throughout the
networks, any observer can access all local hidden variables plus shared
randomness. In the proposed linear and nonlinear Bell-type inequalities, there
are more correlators to reveal genuine Bell locality than those in the
$K$-locality inequalities, and their upper bounds can be specified using the
probability normalization of the predetermined probability distribution. On
the other hand, the no-clone theorem limits the broadcast of quantum
correlations in quantum networks. To explore genuine Bell nonlocality, the
stabilizing operators play an important role in designing the segmented Bell
operators and assigning the incompatible measurements for the spatially
separated observers. We prove the maximal violations of the proposed Bell-type
inequalities tailored for the given qubit distributions in quantum networks.

\end{abstract}
\volumeyear{year}
\volumenumber{number}
\issuenumber{number}
\eid{identifier}
\date[Date text]{date}
\received[Received text]{date}

\revised[Revised text]{date}

\accepted[Accepted text]{date}

\published[Published text]{date}

\startpage{1}
\endpage{2}
\maketitle

\section{Introduction}

As one of the central features in quantum foundations, Bell theorem states
that quantum correlations between spatially separated systems can break the
limits of classical causal relations \cite{1,2}. In ontological models and
local hidden variables (LHV), the locality of spacelike events and the realism
therein constrain the strength of classical correlations, which are
upper-bounded by the Bell inequalities. Quantum theory inconsistent with local
realism predicts stronger correlations that violate Bell inequalities. Thanks
to quantum information science, two-particle and multiparticle quantum
correlations have been extensively investigated. In addition to its immense
influence on quantum foundations, Bell nonlocality as a quantum resource has
also led to applications in quantum information processing such as quantum
cryptography \cite{3}, reductions in communication complexity \cite{4},
private random number generation \cite{5}, and self-testing protocols \cite{6}.

To test the strength of Bell nonlocality, an associated Bell test is devised
for a given Bell inequality. Therein, a source initially emits a state of two
or more particles received by spatially separated observers, who each perform
local measurements with a random measurement setting as the input and then
obtain the measurement outcome as the output. The observers are to verify
whether or not the strength of the input-output correlations violates the
tested Bell inequality. Although there is still a decent chance of generating
Bell-violating correlations using randomly chosen measurement bases
\cite{7,8}, the Bell inequality and local observables in the Bell test should
be deliberately set so that the prepared entangled state can achieve the
maximum violation. In particular, given the prepared entangled state as a
stabilizer state, the stabilizing operators can be used to assign the local
observables and design the associated Bell operators.

As an example, denote the four two-qubit Bell states by $\left\vert \Phi^{\pm
}\right\rangle $ $=\frac{1}{\sqrt{2}}(\left\vert 00\right\rangle \pm\left\vert
11\right\rangle )$ and $\left\vert \Psi^{+}\right\rangle =\frac{1}{\sqrt{2}%
}(\left\vert 01\right\rangle \pm\left\vert 10\right\rangle )$. The operators
$\sigma_{z}\otimes$ $\sigma_{z}$ ($\sigma_{x}\otimes$ $\sigma_{x}$) can
stabilize a two-dimensional subspace spanned by two orthogonal two-qubit Bell
states $\left\vert \Phi^{+}\right\rangle $ and $\left\vert \Phi^{-}%
\right\rangle $ ($\left\vert \Phi^{+}\right\rangle $ and $\left\vert \Psi
^{+}\right\rangle $). The intersection of a two-dimensional subspace
stabilized by $\sigma_{z}\otimes$ $\sigma_{z}$ and $\sigma_{x}\otimes$
$\sigma_{x}$ is the Bell state $\left\vert \Phi^{+}\right\rangle $. Regarding
the Clauser-Horne-Shimony-Holt (CHSH) inequality as the Bell inequality
tailored for $\left\vert \Phi^{+}\right\rangle $, two segmented Bell operators
therein are devised proportional to these two stabilizing operators. In
detail, let the local measurement pairs for separate Alice and Bob in the CHSH
test be $(A_{0}$, $A_{1})$ and $(B_{0}$, $B_{1})$, respectively, and the
Bell-CHSH operator be $\mathbf{B}_{CHSH}=\sum\nolimits_{i,\text{ }j=0}%
^{1}(-1)^{ij}A_{i}\otimes B_{j}$. The CHSH inequality reads $\left\langle
\mathbf{B}_{CHSH}\right\rangle _{LHV}\leq2$, where $\left\langle
\cdot\right\rangle $ denotes the expectation value of $\cdot$. To achieve the
maximum CHSH value, assign the observables $\sigma_{z}\rightarrow\frac
{1}{\sqrt{2}}(A_{0}+A_{1})$ and $\sigma_{x}\rightarrow\frac{1}{\sqrt{2}}%
(A_{0}-A_{1})$ for the first qubit; and assign the observables $\sigma
_{z}\rightarrow B_{0}$, $\sigma_{x}\rightarrow B_{1}$ for the second qubit.
Consequently, the Bell-CHSH operator comprises two segmented Bell operators :
$(A_{0}+A_{1})B_{0}=\sqrt{2}(\sigma_{z}\otimes$ $\sigma_{z})$ and
$(A_{0}-A_{1})$ $B_{1}=\sqrt{2}(\sigma_{x}\otimes$ $\sigma_{x})$ and is
proportional to the addition of two stabilizing operators. As a result, the
maximum CHSH value $2\sqrt{2}$ can be achieved using the Bell state
$\left\vert \Phi^{+}\right\rangle $.

Recently, Bell nonlocality in quantum networks has attracted much research
attention. A quantum network involves multiple independent sources, and each
of them initially emits an entangled state of two or more qubits distributed
among specific observers. Since long-distance quantum networks involving
large-scale multi-users are the essential goals of upcoming quantum
communication, it is fundamental to study their nonlocality strength in the
network. Specifically, the bilocality and $K$-locality correlations in the
two-source and later the $K$-sources cases, respectively, have been
extensively studied \cite{99,12,125,126,acin,Luo,9,127,13,rev}. In most of the
literature, however, the notions of bilocality and $K$-locality implicitly
refer to the restricted distributions of local hidden variables. For example,
in the classical bilocal scenario, a source $e_{1}$ sends the local hidden
variable $\lambda_{1}$ to Alice and Bob, and the other independent separate
source $e_{2}$ sends the local hidden variable $\lambda_{2}$ to Charlie and
Bob. In this case, only Bob can access both $\lambda_{1}$ and $\lambda_{2}$.
Regarding local hidden variables as classical objects that can be perfectly
cloned and then spread throughout the networks, hereafter genuine Bell
locality refers to the correlation strength in the scenario in which all local
hidden variables and shared randomness are accessible to each observer in the
classical networks.

Here, we explore Bell nonlocality in the $K$-source quantum networks. Therein,
each quantum source emits the two-qubit Bell state to spatially separated
observers. An observer can receive more than two or more qubits from different
quantum sources and then perform a joint measurement on these accessible
qubits. The paper is organized as follows. In Sec. II and III, we explore the
two-source networks ($K=2$) and the $K$-source star-networks, respectively. We
propose linear Bell inequalities that respect genuine Bell locality and, on
the other hand, can be maximally violated using the prepared state $\left\vert
\Phi^{+}\right\rangle ^{\otimes K}$ distributed in the quantum networks. In
Sec. IV, we propose nonlinear Bell inequalities tailored for the distributed
product state $\left\vert \Phi^{+}\right\rangle ^{\otimes K}$. In Sec. V, we
investigate the Bell nonlocality in ($N$, $K$, $m$) networks as an extension
of the star-networks. \ Finally, we revisit the two-source quantum networks in
Sec. VI. There, a source allows for the emission of three-qubit
Greenberger-Horne-Zeilinger (GHZ) states instead of the two-qubit state
$\left\vert \Phi^{+}\right\rangle $.\ Since two-qubit Bell states and GHZ
states are stabilizer states, the stabilizing operators are demonstrated to
play a substantial role in constructing the Bell-type inequalities and
assigning local observables.

\section{Linear Bell inequalities in two-source networks}

We review the bilocal model as follows. The source $e_{1}$\ ($e_{2}$)
initially sends two particles involving the local hidden variable $\lambda
_{1}$ ($\lambda_{2}$) to Alice and Bob (Charlie and Bob). Alice, Bob, and
Charles perform measurements on their accessible particles, labeled by $x$,
$y$, and $z$, and obtain outcomes denoted by $a_{x}$, $b_{y}$, and $c_{z}$,
respectively. By bilocality, the\ tripartite distribution can be written in
the factorized form%

\begin{equation}
P(a_{x},b_{y},c_{z}|x,y,z)=%
%TCIMACRO{\dsum \nolimits_{\lambda_{1},\lambda_{2}}}%
%BeginExpansion
{\displaystyle\sum\nolimits_{\lambda_{1},\lambda_{2}}}
%EndExpansion
P(\lambda_{1},\lambda_{2})P(a_{x}|x,\lambda_{1})P(b_{y}|y,\lambda_{1}%
,\lambda_{2})P(c_{z}|z,\lambda_{2}).\label{b1}%
\end{equation}
The response function for Alice depends only on the hidden state $\lambda_{1}%
$, the one for Charlie only on $\lambda_{2}$; while the one for Bob on
$\lambda_{1}$ and $\lambda_{2}$. Furthermore, it is assumed that two sources
$e_{1}$ and $e_{2}$\ are independent and uncorrelated. We denote $\rho
(\lambda_{i})$ the distribution of the local hidden variable $\lambda_{i}$,
and the joint distribution $\rho(\lambda_{1},\lambda_{2})$ can be factorized
as $\rho(\lambda_{1})\rho(\lambda_{2})$ in the bilocal model. A typical
nonlinear bilocal inequality can be characterized as \cite{99,12,125,126}
\begin{equation}
\left\vert \left\langle \mathbf{I}\right\rangle _{BM}\right\vert ^{\frac{1}%
{2}}+\left\vert \left\langle \mathbf{J}\right\rangle _{BM}\right\vert
^{\frac{1}{2}}\leq1.\label{bi}%
\end{equation}
Violating (\ref{bi}) just indicates that the correlation strength in the
network is non-bilocal. It is also claimed that any LHV in the bilocal model
respects the linear Bell inequality%
\begin{equation}
\left\vert \left\langle \mathbf{I}\right\rangle _{LHV}\right\vert +\left\vert
\left\langle \mathbf{J}\right\rangle _{LHV}\right\vert \leq1,\label{bil}%
\end{equation}
which indicates that even classical correlations obeying (\ref{bil}) can admit
bi-nonlocality that violates (\ref{bi}) \cite{99,12,125,126}.

A few remarks on bilocality and non-bilocality are in order. Firstly, to
achieve the maximum quantum values of (\ref{bi}) and (\ref{bil}), the
following prepare-and-measure scenario is usually proposed in most of the
literature. The source $e_{1}$ emits the two-qubit states $\left\vert
e_{1}\right\rangle _{12}=\left\vert \Phi^{+}\right\rangle $ of qubits 1 and 2
sent to Alice and Bob, respectively; the source $e_{2}$ emits the two-qubit
states $\left\vert e_{2}\right\rangle _{34}=\left\vert \Phi^{+}\right\rangle $
of qubits 3 and 4 sent to Bob and Charlie, respectively. Bob can measure the
joint observables $Z_{2}Z_{3}$ or $X_{2}X_{3}$ that are compatible with each
other, or equivalently, perform the Bell-state measurement using the basis
$\{\left\vert \Phi^{\pm}\right\rangle ,\left\vert \Psi^{\pm}\right\rangle \}$.
This Bell-state measurement brings the entanglement swapping and creates
nonlocal correlations between two particles that never interact with each
other \cite{99}. It will be shown that the segmented Bell operators
$\mathbf{I}$ and $\mathbf{J}$ in (\ref{bi}) each are proportional to
stabilizing operators of $\left\vert \Phi^{+}\right\rangle ^{\otimes2}$
\cite{12,13}. Secondly, the post-selected density matrix after local
measurements is also known as the four-qubit bound-entangled Smolin state%

\begin{align}
\rho_{Smolin} &  =\frac{1}{4}([\Phi^{+}]_{12}[\Phi^{+}]_{34}+[\Phi^{-}%
]_{12}[\Phi^{-}]_{34}+[\Psi^{+}]_{12}[\Psi^{+}]_{34}+[\Psi^{-}]_{12}[\Psi
^{-}]_{34})\nonumber\\
&  =\frac{1}{4}([\Phi^{+}]_{23}[\Phi^{+}]_{14}+[\Phi^{-}]_{23}[\Phi^{-}%
]_{14}+[\Psi^{+}]_{23}[\Psi^{+}]_{14}+[\Psi^{-}]_{23}[\Psi^{-}]_{14}%
)\nonumber\\
&  =\frac{1}{4}([\Phi^{+}]_{13}[\Phi^{+}]_{24}+[\Phi^{-}]_{13}[\Phi^{-}%
]_{24}+[\Psi^{+}]_{13}[\Psi^{+}]_{24}+[\Psi^{-}]_{13}[\Psi^{-}]_{24}%
),\label{Smolin}%
\end{align}
where the one-rank projector is denoted by $[x]_{ij}=(\left\vert
x\right\rangle _{ij})(_{ij}\left\langle x\right\vert )$ \cite{smolin}. It is
easy to verify that $\rho_{Smolin}$ as well as $\left\vert \Phi^{+}%
\right\rangle _{12}\left\vert \Phi^{+}\right\rangle _{34}$ can achieve the
maximum quantum value of (\ref{bi}) and (\ref{bil}). \ (\ref{Smolin})
indicates that $\rho_{Smolin}$ is separable across any bipartite cut, and
hence there are several ways to prepare $\rho_{Smolin}$ in quantum networks.
According to the first line of (\ref{Smolin}), $\rho_{Smolin}$ can be prepared
with the help of shared randomness such that $\left\vert e_{1}\right\rangle
_{12}=\left\vert e_{2}\right\rangle _{34}\in\{\left\vert \Phi^{\pm
}\right\rangle $, $\left\vert \Psi^{\pm}\right\rangle \}$ for each emission.
On the other hand, according to the second line of (\ref{Smolin}), let
$\rho_{Smolin}$ be prepared with emitted states $\left\vert e_{1}\right\rangle
_{23}=\left\vert e_{2}\right\rangle _{14}\in\{\left\vert \Phi^{\pm
}\right\rangle $, $\left\vert \Psi^{\pm}\right\rangle \}$. Bob receives qubits
2 and 3 from the same source $e_{1}$ and Alice and Charlie receive qubits 1
and 4, respectively, from the source $e_{2}$. In this case, Bob's Bell-state
measurement does not lead to entanglement swapping but simply obtains two-bit
classical information without any disturbance. Furthermore, these methods of
preparing $\rho_{Smolin}$ show that any particle can share the same hidden
variable with any of the other three in the bilocal model. As a result, the
notions of bilocality and non-bilocality are not operationally meaningful in
the two-source networks.

To guarantee genuine Bell locality in classical two-source networks, it is
assumed that all observers receive the local hidden variable $\lambda
=\lambda_{1}\cup\lambda_{2}$ plus shared randomness. We propose linear and
nonlinear Bell inequalities that are maximally violated by the pure state
$\left\vert \Phi^{+}\right\rangle _{12}\left\vert \Phi^{+}\right\rangle _{34}%
$. \ To do this, note that the Bell operators $\mathbf{I}$ and $\mathbf{J}$ in
(\ref{bi}) will be shown proportional to $Z_{1}Z_{2}Z_{3}Z_{4}$ and
$X_{1}X_{2}X_{3}X_{4}$, respectively, stabilizing both $\left\vert \Phi
^{+}\right\rangle ^{\otimes2}$ and $\rho_{Smolin}$. Two other operators
$Z_{1}Z_{2}X_{3}X_{4}$\ and $X_{1}X_{2}Z_{3}Z_{4}$ stabilizing $\left\vert
\Phi^{+}\right\rangle ^{\otimes2}$ rather than $\rho_{Smolin}$ must be
involved in the Bell inequality. Denote the correlators by $\mathbf{I}%
_{y_{0}y_{1}}=\frac{1}{4}%
%TCIMACRO{\dsum \nolimits_{x,z}}%
%BeginExpansion
{\displaystyle\sum\nolimits_{x,z}}
%EndExpansion
(-1)^{xy_{0}+zy_{1}}a_{x}b_{y_{0}y_{1}}c_{z}$, where $a_{x}$ and $c_{z}$ are
the outcomes of the observables $A_{x}$, $C_{z}$ measured by Alice, Charlie,
respectively;\ $b_{y_{0}y_{1}}$is the product of the outcomes of two local
observables $B_{y_{1}y_{2}}$ and $x$, $y_{0}$, $y_{1}$, $z\in\{0,1\}$ and
$a_{x}$, $b_{y_{0}y_{1}}$, $c_{z}\in\{-1,1\}$. The joint probability in
classical networks is defined as%

\begin{align}
&  p_{y_{0}y_{1}}\nonumber\\
&  =%
%TCIMACRO{\dsum \nolimits_{\lambda}}%
%BeginExpansion
{\displaystyle\sum\nolimits_{\lambda}}
%EndExpansion
P(\left\vert \frac{(a_{0}+(-1)^{y_{0}}a_{1})}{2}\frac{(c_{0}+(-1)^{y_{1}}%
c_{1})}{2}\right\vert =1|\lambda)P(\lambda)\nonumber\\
&  =%
%TCIMACRO{\dsum \nolimits_{\lambda}}%
%BeginExpansion
{\displaystyle\sum\nolimits_{\lambda}}
%EndExpansion
P(\left\vert \frac{(a_{0}+(-1)^{y_{0}}a_{1})}{2}b_{y_{0}y_{1}}\frac
{(c_{0}+(-1)^{y_{1}}c_{1})}{2}\right\vert =1|\lambda)P(\lambda), \label{p1}%
\end{align}
and the probability normalization requires that $%
%TCIMACRO{\dsum \nolimits_{y_{0},y_{1}}}%
%BeginExpansion
{\displaystyle\sum\nolimits_{y_{0},y_{1}}}
%EndExpansion
p_{y_{0}y_{1}}=1$. Note that $p_{y_{0}y_{1}}=\left\langle \left\vert
\mathbf{I}_{y_{0}y_{1}}\right\vert \right\rangle _{LHV}$, and hence the first
proposed linear Bell inequality reads
\begin{align}
&  \left\langle \mathbf{I}_{00}\right\rangle _{LHV}+\left\langle
\mathbf{I}_{01}\right\rangle _{LHV}+\left\langle \mathbf{I}_{10}\right\rangle
_{LHV}+\left\langle \mathbf{I}_{11}\right\rangle _{LHV}\nonumber\\
&  \leq\left\vert \left\langle \mathbf{I}_{00}\right\rangle _{LHV}\right\vert
+\left\vert \left\langle \mathbf{I}_{01}\right\rangle _{LHV}\right\vert
+\left\vert \left\langle \mathbf{I}_{10}\right\rangle _{LHV}\right\vert
+\left\vert \left\langle \mathbf{I}_{11}\right\rangle _{LHV}\right\vert
\nonumber\\
&  \leq\left\langle \left\vert \mathbf{I}_{00}\right\vert \right\rangle
_{LHV}+\left\langle \left\vert \mathbf{I}_{01}\right\vert \right\rangle
_{LHV}+\left\langle \left\vert \mathbf{I}_{10}\right\vert \right\rangle
_{LHV}+\left\langle \left\vert \mathbf{I}_{11}\right\vert \right\rangle
_{LHV}\nonumber\\
&  =%
%TCIMACRO{\dsum \nolimits_{ij}}%
%BeginExpansion
{\displaystyle\sum\nolimits_{ij}}
%EndExpansion
p_{ij}\nonumber\\
&  =1, \label{L}%
\end{align}
where the second inequality is due to the fact $\left\vert \left\langle
\cdot\right\rangle \right\vert $ $\leq\left\langle \left\vert \cdot\right\vert
\right\rangle $. It will be shown that probability normalization plays an
important role in constructing Bell inequalities throughout the paper.

In the quantum version, denote the segmented Bell operators $\mathbf{I}%
_{y_{0}y_{1}}=\frac{1}{4}(A_{0}+(-1)^{y_{0}}A_{1})B_{y_{0}y_{1}}%
(C_{0}+(-1)^{y_{1}}C_{1})$. To achieve the maximal violation of (\ref{L}) in
the two-source quantum networks, assign the local observables $\sigma
_{z}\rightarrow\frac{1}{\sqrt{2}}(A_{0}+A_{1})$, $\sigma_{x}\rightarrow
\frac{1}{\sqrt{2}}(A_{0}-A_{1}),\sigma_{z}\rightarrow\frac{1}{\sqrt{2}}%
(C_{0}+C_{1})$, $\sigma_{x}\rightarrow\frac{1}{\sqrt{2}}(C_{0}-C_{1}%
),\sigma_{z}\sigma_{z}\rightarrow B_{00}$, $\sigma_{x}\sigma_{x}\rightarrow
B_{11},\sigma_{z}\sigma_{x}\rightarrow B_{01}$, $\sigma_{x}\sigma
_{z}\rightarrow B_{10}.$ As a result, we have $\mathbf{I}_{00}\mathbf{=}%
\frac{1}{2}Z_{1}Z_{2}Z_{3}Z_{4}$, $\mathbf{I}_{11}\mathbf{=}\frac{1}{2}%
X_{1}X_{2}X_{3}X_{4}$, $\mathbf{I}_{01}\mathbf{=}\frac{1}{2}Z_{1}Z_{2}%
X_{3}X_{4}$ and $\mathbf{I}_{10}\mathbf{=}\frac{1}{2}X_{1}X_{2}Z_{3}Z_{4}$. We have%

\begin{equation}
\max\{\left\langle \mathbf{I}_{00}\right\rangle _{Q}+\left\langle
\mathbf{I}_{01}\right\rangle _{Q}+\left\langle \mathbf{I}_{10}\right\rangle
_{Q}+\left\langle \mathbf{I}_{11}\right\rangle _{Q}\}=2. \label{1}%
\end{equation}
On the other hand, since $\mathbf{I}_{00}\mathbf{I}_{01}\mathbf{I}_{10}=$
$\mathbf{I}_{11}$, only three of these four operators $\mathbf{I}_{00}$,
$\mathbf{I}_{01}$, $\mathbf{I}_{10}$, and $\mathbf{I}_{11}$ are independent
and can stabilize the two-dimensional subspace spanned by $\left\vert \Phi
^{+}\right\rangle _{12}\left\vert \Phi^{+}\right\rangle _{34}$ and $\left\vert
\Psi^{-}\right\rangle _{12}\left\vert \Psi^{-}\right\rangle _{34}$. That is,
the maximum value in (\ref{1}) can be achieved using the mixed state $\rho
_{1}=q_{1}\left[  \Phi^{+}\right]  _{12}\left[  \Phi^{+}\right]
_{34}+(1-q_{1})\left[  \Psi^{-}\right]  _{12}\left[  \Psi^{-}\right]  _{34}$.

Here, we propose the second linear Bell inequality similar to (\ref{L}).
Denote the correlator $\mathbf{I}_{y_{0}^{\prime}y_{1}^{\prime}}^{\prime
}=\frac{1}{4}%
%TCIMACRO{\dsum \nolimits_{x,z}}%
%BeginExpansion
{\displaystyle\sum\nolimits_{x,z}}
%EndExpansion
(-1)^{x^{\prime}y_{0}^{\prime}+z^{\prime}y_{1}^{\prime}}a_{x^{\prime}}%
^{\prime}b_{y_{0}^{\prime}y_{1}^{\prime}}^{\prime}c_{z^{\prime}}^{\prime}$,
where $a_{x}^{\prime}$ and $c_{z}^{\prime}$, $b_{y_{0}y_{1}}^{\prime}$ are the
outcomes of the observables $A_{x^{\prime}}^{\prime}$, $C_{z^{\prime}}%
^{\prime}$ and $B_{y_{0}^{\prime}y_{1}^{\prime}}^{\prime}$ measured by Alice,
Charlie and Bob, respectively, and $x^{\prime}$, $y_{0}^{\prime}$,
$y_{1}^{\prime}$, $z^{\prime}\in\{0,1\}$ and $a_{x^{\prime}}^{\prime}$,
$b_{y_{0}^{\prime}y_{1}^{\prime}}^{\prime}$, $c_{z}^{\prime}\in\{-1,1\}$.
Another joint probability in classical networks is defined as%

\begin{align}
&  p_{y_{0}^{\prime}y_{1}^{\prime}}^{\prime}\nonumber\\
&  =%
%TCIMACRO{\dsum \nolimits_{\lambda}}%
%BeginExpansion
{\displaystyle\sum\nolimits_{\lambda}}
%EndExpansion
P(\left\vert \frac{(a_{0}^{\prime}+(-1)^{y_{0}^{\prime}}a_{1}^{\prime})}%
{2}\frac{(c_{0}^{\prime}+(-1)^{y_{1}^{\prime}}c_{1}^{\prime})}{2}\right\vert
=1|\lambda)P(\lambda)\nonumber\\
&  =%
%TCIMACRO{\dsum \nolimits_{\lambda}}%
%BeginExpansion
{\displaystyle\sum\nolimits_{\lambda}}
%EndExpansion
P(\left\vert \frac{(a_{0}^{\prime}+(-1)^{y_{0}^{\prime}}a_{1}^{\prime})}%
{2}b_{y_{0}^{\prime}y_{1}^{\prime}}^{\prime}\frac{(c_{0}^{\prime}%
+(-1)^{y_{1}^{\prime}}c_{1}^{\prime})}{2}\right\vert =1|\lambda)P(\lambda),
\end{align}
and the probability normalization requires that $%
%TCIMACRO{\dsum \nolimits_{y_{0}^{\prime},y_{1}^{\prime}}}%
%BeginExpansion
{\displaystyle\sum\nolimits_{y_{0}^{\prime},y_{1}^{\prime}}}
%EndExpansion
p_{y_{0}^{\prime}y_{1}^{\prime}}^{\prime}=1$. Note that $p_{y_{0}^{\prime
}y_{1}^{\prime}}^{\prime}=\left\langle \left\vert I_{y_{0}^{\prime}%
y_{1}^{\prime}}^{\prime}\right\vert \right\rangle _{LHV}$, and hence the
second linear Bell-type inequality reads
\begin{align}
&  \left\langle \mathbf{I}_{00}^{\prime}\right\rangle _{LHV}+\left\langle
\mathbf{I}_{11}^{\prime}\right\rangle _{LHV}-\left\langle \mathbf{I}%
_{01}^{\prime}\right\rangle _{LHV}-\left\langle \mathbf{I}_{10}^{\prime
}\right\rangle _{LHV}\nonumber\\
&  \leq\left\vert \left\langle \mathbf{I}_{00}^{\prime}\right\rangle
_{LHV}\right\vert +\left\vert \left\langle \mathbf{I}_{11}^{\prime
}\right\rangle _{LHV}\right\vert +\left\vert \left\langle \mathbf{I}%
_{01}^{\prime}\right\rangle _{LHV}\right\vert +\left\vert \left\langle
\mathbf{I}_{10}^{\prime}\right\rangle _{LHV}\right\vert \nonumber\\
&  \leq\left\langle \left\vert \mathbf{I}_{00}^{\prime}\right\vert
\right\rangle _{LHV}+\left\langle \left\vert \mathbf{I}_{11}^{\prime
}\right\vert \right\rangle _{LHV}+\left\langle \left\vert \mathbf{I}%
_{01}^{\prime}\right\vert \right\rangle _{LHV}+\left\langle \left\vert
\mathbf{I}_{10}^{\prime}\right\vert \right\rangle _{LHV}\nonumber\\
&  =%
%TCIMACRO{\dsum \nolimits_{ij}}%
%BeginExpansion
{\displaystyle\sum\nolimits_{ij}}
%EndExpansion
p_{y_{0}y_{1}}^{\prime}\nonumber\\
&  =1, \label{L11}%
\end{align}

In the quantum region, the key is to take advantage of the stabilizing
operator $-\sigma_{y}\otimes$ $\sigma_{y}=(\sigma_{z}\otimes$ $\sigma
_{z})(\sigma_{x}\otimes$ $\sigma_{x})$ of $\left\vert \Phi^{+}\right\rangle $.
Denote the segmented Bell operators\emph{\ }$\mathbf{I}_{y_{0}^{\prime}%
y_{1}^{\prime}}^{\prime}=\frac{1}{4}(A_{0}^{\prime}+(-1)^{y_{0}^{\prime}}%
A_{1}^{\prime})B_{y_{0}y_{1}}^{\prime}(C_{0}^{\prime}+(-1)^{y_{1}^{\prime}%
}C_{1}^{\prime})$ and set the observables $\sigma_{z}\rightarrow\frac{1}%
{\sqrt{2}}(A_{0}^{\prime}+A_{1}^{\prime})$, $\sigma_{y}\rightarrow\frac
{1}{\sqrt{2}}(A_{0}^{\prime}-A_{1}^{\prime}),\sigma_{z}\rightarrow\frac
{1}{\sqrt{2}}(C_{0}^{\prime}+C_{1}^{\prime})$, $\sigma_{y}\rightarrow\frac
{1}{\sqrt{2}}(C_{0}^{\prime}-C_{1}^{\prime}),\sigma_{z}\sigma_{z}\rightarrow
B_{00}^{\prime}=B_{00}$, $\sigma_{y}\sigma_{y}\rightarrow B_{11}^{\prime
},\sigma_{z}\sigma_{y}\rightarrow B_{01}^{\prime}$, and $\sigma_{y}\sigma
_{z}\rightarrow B_{10}^{\prime}.$As a result, $\mathbf{I}_{00}^{\prime
}=\mathbf{I}_{00}$, $\mathbf{I}_{11}^{\prime}\mathbf{=}\frac{1}{2}Y_{1}%
Y_{2}Y_{3}Y_{4}$, $\mathbf{I}_{01}^{\prime}\mathbf{=}\frac{1}{2}Z_{1}%
Z_{2}Y_{3}Y_{4}$ and $\mathbf{I}_{10}^{\prime}\mathbf{=}\frac{1}{2}Y_{1}%
Y_{2}Z_{3}Z_{4}$, where $\mathbf{I}_{ij}^{\prime}$ can be also obtained by
replacing the Pauli observable $X_{k}$ in $\mathbf{I}_{ij}$\ by $iY_{k}$. In
this case, we have
\begin{align}
&  \max\{\left\langle \mathbf{I}_{00}^{\prime}\right\rangle _{Q}+\left\langle
\mathbf{I}_{11}^{\prime}\right\rangle _{Q}-\left\langle \mathbf{I}%
_{01}^{\prime}\right\rangle _{Q}-\left\langle \mathbf{I}_{10}^{\prime
}\right\rangle _{Q}\}\nonumber\\
&  =\max\{\left\vert \left\langle \mathbf{I}_{00}^{\prime}\right\rangle
_{Q}\right\vert +\left\vert \left\langle \mathbf{I}_{11}^{\prime}\right\rangle
_{Q}\right\vert +\left\vert \left\langle \mathbf{I}_{01}^{\prime}\right\rangle
_{Q}\right\vert +\left\vert \left\langle \mathbf{I}_{10}^{\prime}\right\rangle
_{Q}\right\vert \}\nonumber\\
&  =2. \label{L2}%
\end{align}
Similarly, since $\mathbf{I}_{00}^{\prime}\mathbf{I}_{01}^{\prime}%
\mathbf{I}_{10}^{\prime}=$ $\mathbf{I}_{11}^{\prime}$, only three of these
four operators $\mathbf{I}_{00}^{\prime}$, $\mathbf{I}_{01}^{\prime}$,
$\mathbf{I}_{10}^{\prime}$, and $\mathbf{I}_{11}^{\prime}$ are independent and
can stabilize a two-dimensional subspace spanned by $\left\vert \Phi
^{+}\right\rangle _{12}\left\vert \Phi^{+}\right\rangle _{34}$ and $\left\vert
\Psi^{+}\right\rangle _{12}\left\vert \Psi^{+}\right\rangle _{34}$ . That is,
the maximum quanutm value (\ref{L2}) can be achieved using the state $\rho
_{2}=q_{2}\left[  \Phi^{+}\right]  _{12}\left[  \Phi^{+}\right]
_{34}+(1-q_{2})\left[  \Psi^{+}\right]  _{12}\left[  \Psi^{+}\right]  _{34}$.
Finally, to achieve (\ref{L}) and (\ref{L2}) simultaneously, the relation
$\rho_{1}$ $=\rho_{2}=\left[  \Phi^{+}\right]  \left[  \Phi^{+}\right]  $ with
$q_{1}=q_{2}=1$ must hold.

A few remarks are in order. Denote the Bell states $\left\vert \phi^{\pm
}\right\rangle =I_{2}\otimes H\left\vert \Phi^{\pm}\right\rangle $,
$\left\vert \psi^{\pm}\right\rangle =I_{2}\otimes H\left\vert \Psi^{\pm
}\right\rangle $, $\left\vert \phi^{\prime\pm}\right\rangle =I_{2}\otimes
H^{\prime}\left\vert \Phi^{\pm}\right\rangle $, $\left\vert \psi^{\pm
}\right\rangle =I_{2}\otimes H^{\prime}\left\vert \Psi^{\pm}\right\rangle $,
where
\[
I_{2}=\left(
\begin{array}
[c]{cc}%
1 & 0\\
0 & 1
\end{array}
\right)  \text{, }H=\frac{1}{\sqrt{2}}\left(
\begin{array}
[c]{cc}%
1 & 1\\
1 & -1
\end{array}
\right)  ,\text{ }H^{\prime}=\frac{1}{\sqrt{2}}\left(
\begin{array}
[c]{cc}%
1 & -i\\
i & -1
\end{array}
\right)  \text{.}%
\]
In the Bell test associated with the Bell inequality (\ref{L}), Bob can
perform the Bell-state measurement either with the basis $\{\left\vert
\Phi^{\pm}\right\rangle ,\left\vert \Psi^{\pm}\right\rangle \}$ instead of the
joint measurements $\sigma_{z}\sigma_{z}$ and $\sigma_{x}\sigma_{x}$ or with
the basis $\{\left\vert \phi^{\pm}\right\rangle ,\left\vert \psi^{\pm
}\right\rangle \}$ instead of the joint measurements $\sigma_{z}\sigma_{x}$
and $\sigma_{x}\sigma_{z}$; in the Bell test associated with the Bell
inequality (\ref{L11}), he can perform the Bell-state measurement either with
the basis $\{\left\vert \Phi^{\pm}\right\rangle ,\left\vert \Psi^{\pm
}\right\rangle \}$ instead of the joint measurements $\sigma_{z}\sigma_{z}$
and $\sigma_{y}\sigma_{y}$ or with the basis $\{\left\vert \phi^{\prime\pm
}\right\rangle ,\left\vert \psi^{\prime\pm}\right\rangle \}$ instead of the
joint measurements $\sigma_{z}\sigma_{y}$ and $\sigma_{y}\sigma_{z}$. In
either Bell test, Bob exploits two incompatible Bell-state basis, whereas only
one Bell-state base is used5 in the bilocal Bell test.

In addition, we combine the proposed Bell inequalities (\ref{L}) and
(\ref{L11}) into one that reads%
\[%
%TCIMACRO{\dsum \nolimits_{y_{0},y_{1}}}%
%BeginExpansion
{\displaystyle\sum\nolimits_{y_{0},y_{1}}}
%EndExpansion
(\left\langle \mathbf{I}_{y_{0}y_{1}}\right\rangle +(-1)^{y_{0}+y_{1}%
}\left\langle \mathbf{I}_{y_{0}y_{1}}^{\prime}\right\rangle )\overset
{LHV}{\leq2}.
\]
There are four possible measurement settings for Alice or Charlie. In this
case, the one-bit inputs $x$, $x^{\prime}$, $z$, and $z^{\prime}$ should be
revised as the two-bit ones $0x$, $1x^{\prime}$, $0z$, and $1z^{\prime}$,
respectively. There are seven different local observables for Bob. The two-bit
inputs $y_{0}y_{1}$ and $y_{0}^{\prime}y_{1}^{\prime}$ should be revised as
three-bit inputs $0y_{0}y_{1}$ and $1y_{0}^{\prime}y_{1}^{\prime}$, respectively.

\section{Linear Bell inequalities for $K$-source star-networks}

In the $K$-source star-networks, the source $e_{i}$ emits the particles $i$
and $(K+i)$ that are sent to Bob and Alice$_{i}$, respectively, $1\leq i\leq
K$. Similarly, we construct two linear Bell inequalities. Denote Bob's $K$-bit
input bit string by $y=y_{1}y_{2}\cdots y_{K}$ with the Hamming weight $W(y)$
and the one-bit output by $b_{y}$; Alice$_{i}$ 's input bit by $x_{i}\ $and
output by $a_{x_{i}}^{(i)}$, where the input bits $x_{1}$, $x_{2}$,$\cdots$,
$x_{K}$, $y_{1}$, $y_{2}$,$\cdots$, $y_{K}$ $\in\{0$, $1\}$ and the
measurement outcomes $b_{y}$, $a_{x_{1}}^{(1)}$, $...$, $a_{x_{K}}^{(K)}$
$\in\{1$, $-1\}$. Regarding the genuine Bell locality, Alice$_{1}$,
Alice$_{2}$, \ldots, Alice$_{K}$ and Bob share randomness, and the local
hidden variable $\lambda=\lambda_{1}\cup\lambda_{2}\cdots\cup\lambda_{K}$,
where $\lambda_{i}$ is the local hidden variable emitted from the $i$-th
classical source. Note that one and only one of the pre-assigned values among
these $2^{K}$ variables$\ \left\vert \frac{1}{2^{K}}%
%TCIMACRO{\dprod \nolimits_{i=1}^{K}}%
%BeginExpansion
{\displaystyle\prod\nolimits_{i=1}^{K}}
%EndExpansion
(a_{0}^{(i)}+(-1)^{y_{i}}a_{1}^{(i)})b_{y}\right\vert $ is 1 and all the
others are 0 since the bits $b_{y}$, $a_{x_{1}}^{(1)}$, $...$, $a_{x_{K}%
}^{(K)}$ are predetermined in classical causal models. Denote the correlator
$\mathbf{I}_{y}=\frac{1}{2^{K}}%
%TCIMACRO{\dprod \nolimits_{i=1}^{K}}%
%BeginExpansion
{\displaystyle\prod\nolimits_{i=1}^{K}}
%EndExpansion
(a_{0}^{(i)}+(-1)^{y_{i}}a_{1}^{(i)})b_{y}$ and the probability distribution%

\begin{align}
&  p_{y}\nonumber\\
&  =%
%TCIMACRO{\dsum \nolimits_{\lambda}}%
%BeginExpansion
{\displaystyle\sum\nolimits_{\lambda}}
%EndExpansion
P(\left\vert \frac{1}{2^{K}}%
%TCIMACRO{\dprod \nolimits_{i=1}^{K}}%
%BeginExpansion
{\displaystyle\prod\nolimits_{i=1}^{K}}
%EndExpansion
(a_{0}^{(i)}+(-1)^{y_{i}}a_{1}^{(i)})\right\vert =1|\lambda)P(\lambda
)\nonumber\\
&  =%
%TCIMACRO{\dsum \nolimits_{\lambda}}%
%BeginExpansion
{\displaystyle\sum\nolimits_{\lambda}}
%EndExpansion
P(\left\vert \frac{1}{2^{K}}%
%TCIMACRO{\dprod \nolimits_{i=1}^{K}}%
%BeginExpansion
{\displaystyle\prod\nolimits_{i=1}^{K}}
%EndExpansion
(a_{0}^{(i)}+(-1)^{y_{i}}a_{1}^{(i)})b_{y}\right\vert =1|\lambda
)P(\lambda),\label{p2}%
\end{align}
and the probability nomalization requires that $%
%TCIMACRO{\dsum \nolimits_{y}}%
%BeginExpansion
{\displaystyle\sum\nolimits_{y}}
%EndExpansion
p_{y}=1$. Note that $p_{y}=\left\langle \left\vert \mathbf{I}_{y}\right\vert
\right\rangle _{LHV}$, and the first linear Bell inequality reads
\begin{align}
&
%TCIMACRO{\dsum \nolimits_{y}}%
%BeginExpansion
{\displaystyle\sum\nolimits_{y}}
%EndExpansion
\left\langle \mathbf{I}_{y}\right\rangle _{LHV}\nonumber\\
&  \leq%
%TCIMACRO{\dsum \nolimits_{y}}%
%BeginExpansion
{\displaystyle\sum\nolimits_{y}}
%EndExpansion
\left\langle \left\vert \mathbf{I}_{y}\right\vert \right\rangle _{LHV}%
\nonumber\\
&  =%
%TCIMACRO{\dsum \nolimits_{y}}%
%BeginExpansion
{\displaystyle\sum\nolimits_{y}}
%EndExpansion
p_{y}=1.\label{7}%
\end{align}
In the quantum region, Alice$_{i}$ and Bob measure the observables $A_{x_{i}%
}^{(i)}$ and $B_{y}=%
%TCIMACRO{\dprod \nolimits_{j=1}^{K}}%
%BeginExpansion
{\displaystyle\prod\nolimits_{j=1}^{K}}
%EndExpansion
B_{y_{j}}^{(j)}$, respectively. By assigning the local observable $\sigma
_{z}\rightarrow\frac{1}{\sqrt{2}}(A_{0}^{(i)}+A_{1}^{(i)})$, $\sigma
_{x}\rightarrow\frac{1}{\sqrt{2}}(A_{0}^{(i)}-A_{1}^{(i)})$, and $B_{y_{j}%
}^{(j)}=Z_{i}^{\overline{y_{j}}}X_{i}^{y_{j}}$, the segmented Bell operator by
$\mathbf{I}_{y}=\frac{1}{2^{K}}%
%TCIMACRO{\dprod \nolimits_{i=1}^{K}}%
%BeginExpansion
{\displaystyle\prod\nolimits_{i=1}^{K}}
%EndExpansion
(A_{0}^{(i)}+(-1)^{y_{i}}A_{1}^{(i)})B_{y}$ can be re-expresse as
\begin{equation}
\mathbf{I}_{y}=\frac{1}{\sqrt{2^{K}}}%
%TCIMACRO{\dprod \nolimits_{i=1}^{K}}%
%BeginExpansion
{\displaystyle\prod\nolimits_{i=1}^{K}}
%EndExpansion
(Z_{i}Z_{K+i})^{\overline{y_{i}}}(X_{i}X_{K+i})^{y_{i}}\text{. }%
\end{equation}
Since $Z_{i}Z_{K+i}\left\vert \Phi^{+}\right\rangle _{i(K+i)}=X_{i}%
X_{K+i}\left\vert \Phi^{+}\right\rangle _{i(K+i)}=\left\vert \Phi
^{+}\right\rangle _{i(K+i)}$, it is easy to verify that $\sqrt{2^{K}}I_{y}$ is
a stabilizing operator of the quantum state $%
%TCIMACRO{\dprod \nolimits_{i=1}^{K}}%
%BeginExpansion
{\displaystyle\prod\nolimits_{i=1}^{K}}
%EndExpansion
\left\vert \Phi^{+}\right\rangle _{i(K+i)}$. As a result, we have
\begin{equation}
\max\{%
%TCIMACRO{\dsum \nolimits_{y}}%
%BeginExpansion
{\displaystyle\sum\nolimits_{y}}
%EndExpansion
\left\langle \mathbf{I}_{y}\right\rangle _{Q}\}=\sqrt{2^{K}}.\label{Line1}%
\end{equation}
Although it is very complicated to find out the subspace stabilized by these
$2^{K}$ operators $\sqrt{2^{K}}I_{0...0},\cdots,$ and $\sqrt{2^{K}}I_{1...1}$,
at least $K$ stabilizing operators $\sqrt{2^{K}}I_{y}$ with $W(y)=1$ are
linearly independent. Since there are $2K$ independent stabilizing operators
for the product states $%
%TCIMACRO{\dprod \nolimits_{i=1}^{K}}%
%BeginExpansion
{\displaystyle\prod\nolimits_{i=1}^{K}}
%EndExpansion
\left\vert \Phi^{+}\right\rangle _{i(K+i)}$, we propose the second linear Bell
inequality so that at least another $K$ linearly-independent\ stabilizing
operators are exploited\ as the segmented Bell operators.

Similarly, denote Bob's $K$-bit input string by $y^{\prime}=y_{1}^{\prime
}y_{2}^{\prime}\cdots y_{K}^{\prime}$ and the outcome by $b_{y^{\prime}%
}^{\prime}$; Alice$_{i}$ 's input bit by $x_{i}^{\prime}\ $and outcome by
$a_{x_{i}}^{\prime(i)}$, where input bits $x_{1}^{\prime}$, $x_{2}^{\prime}%
$,$\cdots$, $x_{K}^{\prime}$, $y_{1}^{\prime}$, $y_{2}^{\prime}$,$\cdots$,
$y_{K}^{\prime}$ $\in\{0$, $1\}$ and output bits $b_{y^{\prime}}^{\prime}$,
$a_{x_{1}}^{\prime(1)}$, $...$, $a_{x_{K}}^{\prime(K)}$ $\in\{1$, $-1\}$.
Denote the correlator $\mathbf{I}_{y^{\prime}}^{\prime}=\frac{1}{2^{K}}%
%TCIMACRO{\dprod \nolimits_{i=1}^{K}}%
%BeginExpansion
{\displaystyle\prod\nolimits_{i=1}^{K}}
%EndExpansion
(a_{0}^{\prime(i)}+(-1)^{y_{i}^{\prime}}a_{1}^{\prime(i)})b_{y^{\prime}}$ and
the probability distribution%

\begin{align}
&  p_{y^{\prime}}^{\prime}\nonumber\\
&  =%
%TCIMACRO{\dsum \nolimits_{\lambda}}%
%BeginExpansion
{\displaystyle\sum\nolimits_{\lambda}}
%EndExpansion
P(\left\vert \frac{1}{2^{K}}%
%TCIMACRO{\dprod \nolimits_{i=1}^{K}}%
%BeginExpansion
{\displaystyle\prod\nolimits_{i=1}^{K}}
%EndExpansion
(a_{0}^{\prime(i)}+(-1)^{y_{i}^{\prime}}a_{1}^{\prime(i)})\right\vert
=1|\lambda)P(\lambda)\nonumber\\
&  =%
%TCIMACRO{\dsum \nolimits_{\lambda}}%
%BeginExpansion
{\displaystyle\sum\nolimits_{\lambda}}
%EndExpansion
P(\left\vert \frac{1}{2^{K}}%
%TCIMACRO{\dprod \nolimits_{i=1}^{K}}%
%BeginExpansion
{\displaystyle\prod\nolimits_{i=1}^{K}}
%EndExpansion
(a_{0}^{\prime(i)}+(-1)^{y_{i}^{\prime}}a_{1}^{\prime(i)})b_{y}^{\prime
}\right\vert =1|\lambda)P(\lambda),
\end{align}
According the probability normalization $%
%TCIMACRO{\dsum \nolimits_{y}}%
%BeginExpansion
{\displaystyle\sum\nolimits_{y}}
%EndExpansion
p_{y^{\prime}}^{\prime}=1$and the identity $p_{y^{\prime}}^{\prime
}=\left\langle \left\vert \mathbf{I}_{y}^{\prime}\right\vert \right\rangle
_{LHV}$, the second linear Bell inequality reads
\begin{align}
&
%TCIMACRO{\dsum \nolimits_{y^{\prime}}}%
%BeginExpansion
{\displaystyle\sum\nolimits_{y^{\prime}}}
%EndExpansion
(-1)^{W(y^{\prime})}\left\langle \mathbf{I}_{y^{\prime}}^{\prime}\right\rangle
_{LHV}\nonumber\\
&  \leq%
%TCIMACRO{\dsum \nolimits_{y^{\prime}}}%
%BeginExpansion
{\displaystyle\sum\nolimits_{y^{\prime}}}
%EndExpansion
\left\langle \left\vert \mathbf{I}_{y^{\prime}}^{\prime}\right\vert
\right\rangle _{LHV}\nonumber\\
&  =%
%TCIMACRO{\dsum \nolimits_{y}}%
%BeginExpansion
{\displaystyle\sum\nolimits_{y}}
%EndExpansion
p_{y^{\prime}}^{\prime}=1. \label{8}%
\end{align}
In the quantum region, Alice$_{i}$ and Bob measure the observables
$A_{x_{i}^{\prime}}^{\prime(i)}$ and $B_{y^{\prime}}^{\prime}=%
%TCIMACRO{\dprod \nolimits_{j=1}^{K}}%
%BeginExpansion
{\displaystyle\prod\nolimits_{j=1}^{K}}
%EndExpansion
B_{y_{j}^{\prime}}^{\prime(j)}$, respectively. By assigning the local
observable $\sigma_{z}\rightarrow\frac{1}{\sqrt{2}}(A_{0}^{\prime(i)}%
+A_{1}^{\prime(i)})$, $\sigma_{y}\rightarrow\frac{1}{\sqrt{2}}(A_{0}%
^{\prime(i)}-A_{1}^{\prime(i)}),$ and $B_{y_{j}^{\prime}}^{\prime(j)}%
=Z_{j}^{\overline{y_{i}^{\prime}}}Y_{j}^{y_{i}^{\prime}}$, the segmented Bell
operator denoted by $\mathbf{I}_{y^{\prime}}^{\prime}=\frac{1}{2^{K}}%
%TCIMACRO{\dprod \nolimits_{i=1}^{K}}%
%BeginExpansion
{\displaystyle\prod\nolimits_{i=1}^{K}}
%EndExpansion
(A_{0}^{\prime(i)}+(-1)^{y_{i}^{\prime}}A_{1}^{(i)})B_{y^{\prime}}^{\prime}$
can be re-expresse as
\begin{equation}
\mathbf{I}_{y^{\prime}}^{\prime}=\frac{1}{\sqrt{2^{K}}}%
%TCIMACRO{\dprod \nolimits_{i=1}^{K}}%
%BeginExpansion
{\displaystyle\prod\nolimits_{i=1}^{K}}
%EndExpansion
(Z_{i}Z_{K+i})^{\overline{y_{i}^{\prime}}}(Y_{i}Y_{K+i})^{y_{i}^{\prime}%
}\text{. }%
\end{equation}
It is easy to verify that $\sqrt{2^{K}}(-1)^{W(y^{\prime})}\mathbf{I}%
_{y^{\prime}}^{\prime}$ is a stabilizing operator of the quantum state $%
%TCIMACRO{\dprod \nolimits_{i=1}^{K}}%
%BeginExpansion
{\displaystyle\prod\nolimits_{i=1}^{K}}
%EndExpansion
\left\vert \Phi^{+}\right\rangle _{i(K+i)}$, and we have
\begin{equation}
\max\{%
%TCIMACRO{\dsum \nolimits_{y^{\prime}}}%
%BeginExpansion
{\displaystyle\sum\nolimits_{y^{\prime}}}
%EndExpansion
(-1)^{W(y^{\prime})}\left\langle \mathbf{I}_{y}\right\rangle _{Q}%
\}=\sqrt{2^{K}}. \label{Line2}%
\end{equation}
Therein, $K$ operators $\sqrt{2^{K}}(-1)^{W(y^{\prime})}\mathbf{I}_{y^{\prime
}}^{\prime}$ with $W(y^{\prime})=1$ are linearly indepedent. In the end, these
two linear Bell inequalities (\ref{7}) and (\ref{8}) can be combined into one
that reads%
\begin{equation}%
%TCIMACRO{\dsum \nolimits_{y}}%
%BeginExpansion
{\displaystyle\sum\nolimits_{y}}
%EndExpansion
(\left\langle \mathbf{I}_{y}\right\rangle +(-1)^{W(y)}\left\langle
\mathbf{I}_{y}^{\prime}\right\rangle )\overset{LHV}{\leq}2.
\end{equation}
In the quantum region, we have%

\begin{equation}
\max\{%
%TCIMACRO{\dsum \nolimits_{y}}%
%BeginExpansion
{\displaystyle\sum\nolimits_{y}}
%EndExpansion
(\left\langle \mathbf{I}_{y}\right\rangle _{Q}+(-1)^{W(y)}\left\langle
\mathbf{I}_{y}^{\prime}\right\rangle _{Q})\}=2\sqrt{2^{K}},\label{LinMax}%
\end{equation}
which can be achieved only by the product state $%
%TCIMACRO{\dprod \nolimits_{i=1}^{K}}%
%BeginExpansion
{\displaystyle\prod\nolimits_{i=1}^{K}}
%EndExpansion
\left\vert \Phi^{+}\right\rangle _{i(K+i)}$. The input bits and strings $x$,
$x^{\prime}$, $z$, $z^{\prime}$, $y$, and $y^{\prime}$ in the above two Bell
tests should be revised as $0x$, $1x^{\prime}$, $0z$, $1z^{\prime}$, $0y$, and
$1y^{\prime}$, respectively.

At the end of the section, we prove (\ref{LinMax}) as follows. Without loss of
generality, assign Alice$_{i}$'s local observables as $Z_{K+i}\rightarrow
\frac{1}{2\cos\theta_{i}}(A_{0}^{(i)}+A_{1}^{(i)})$, $X_{K+i}\rightarrow
\frac{1}{2\sin\theta_{i}}(A_{0}^{(i)}-A_{1}^{(i)}),$where $0<\theta_{i}%
<\frac{\pi}{2}$ and $i=1$, $2$, $...$, $K$; Bob's local observables are
assigned as $B_{y}=%
%TCIMACRO{\dprod \nolimits_{i=1}^{K}}%
%BeginExpansion
{\displaystyle\prod\nolimits_{i=1}^{K}}
%EndExpansion
Z_{i}^{\overline{y_{i}}}X_{i}^{y_{i}}$, where the input $K$-bit string
$y=y_{1}y_{2}\cdots y_{K}$. As a result, the segmented Bell operator reads
\begin{equation}
\mathbf{I}_{y}=%
%TCIMACRO{\dprod \nolimits_{i=1}^{K}}%
%BeginExpansion
{\displaystyle\prod\nolimits_{i=1}^{K}}
%EndExpansion
(\cos\theta_{i}Z_{i}Z_{K+i})^{\overline{y_{i}}}(\sin\theta_{i}X_{i}%
X_{K+i})^{y_{i}}\text{. }\label{iy}%
\end{equation}
Since the operators $Z_{i}Z_{K+i}$ and $X_{i}X_{K+i}$ each stabilize the state
$\left\vert e_{i}\right\rangle _{i(K+i)}$, given the product state $%
%TCIMACRO{\dprod \nolimits_{i=1}^{K}}%
%BeginExpansion
{\displaystyle\prod\nolimits_{i=1}^{K}}
%EndExpansion
\left\vert e_{i}\right\rangle _{i(K+i)}$ distributed in the $K$-source quantum
networks, we have
\begin{equation}
\left\langle \mathbf{I}_{y}\right\rangle _{Q}=%
%TCIMACRO{\dprod \nolimits_{i=1}^{K}}%
%BeginExpansion
{\displaystyle\prod\nolimits_{i=1}^{K}}
%EndExpansion
(\cos\theta_{i})^{\overline{y_{i}}}(\sin\theta_{i})^{y_{i}}\text{,}\label{n}%
\end{equation}
and then
\begin{align}
&  \max%
%TCIMACRO{\dsum \nolimits_{y}}%
%BeginExpansion
{\displaystyle\sum\nolimits_{y}}
%EndExpansion
\left\langle \mathbf{I}_{y}\right\rangle _{Q}\nonumber\\
&  =\underset{\left\{  \theta_{i}\right\}  }{\max}%
%TCIMACRO{\dprod \nolimits_{i=1}^{K}}%
%BeginExpansion
{\displaystyle\prod\nolimits_{i=1}^{K}}
%EndExpansion
(\cos\theta_{i}+\sin\theta_{i})\nonumber\\
&  =\underset{\left\{  \theta_{i}\right\}  }{\max}(\sqrt{2})^{K}%
%TCIMACRO{\dprod \nolimits_{i=1}^{K}}%
%BeginExpansion
{\displaystyle\prod\nolimits_{i=1}^{K}}
%EndExpansion
\sin(\theta_{i}+\frac{\pi}{4})\nonumber\\
&  =2^{\frac{K}{2}},\label{max}%
\end{align}
where the equality holds if $\theta_{1}=\theta_{2}=\cdots=\theta_{K}=\frac
{\pi}{4}$. Similarly, assign the observables
\[
Z_{K+i}\rightarrow\frac{1}{2\cos\theta_{i}^{\prime}}(A_{0}^{\prime(i)}%
+A_{1}^{\prime(i)})\text{, }Y_{K+i}\rightarrow\frac{1}{2\sin\theta_{i}%
^{\prime}}(A_{0}^{\prime(i)}-A_{1}^{\prime(i)}),
\]
we have $\max%
%TCIMACRO{\dsum \nolimits_{y}}%
%BeginExpansion
{\displaystyle\sum\nolimits_{y}}
%EndExpansion
(-1)^{W(y)}\left\langle \mathbf{I}_{y}^{\prime}\right\rangle _{Q}=2^{\frac
{K}{2}}$. Therefore, we prove the maximum violation (\ref{LinMax}). Finally,
combining the result (\ref{1}) and (\ref{L2}), we reach (\ref{LinMax}) in the
case $K=2$.

\section{Non-linear Bell-type inequalities for $K$-source star-network}

The nonlinear Bell inequalities in the $K$\textit{-}source ($K\geq2$)\textit{
}classical star\ networks reads
\begin{align}
&
%TCIMACRO{\dsum \nolimits_{y}}%
%BeginExpansion
{\displaystyle\sum\nolimits_{y}}
%EndExpansion
\left\langle \mathbf{I}_{y}\right\rangle _{LHV}^{r}\nonumber\\
&  \leq%
%TCIMACRO{\dsum \nolimits_{y}}%
%BeginExpansion
{\displaystyle\sum\nolimits_{y}}
%EndExpansion
\left\langle \left\vert \mathbf{I}_{y}\right\vert \right\rangle _{LHV}%
^{r}\nonumber\\
&  =%
%TCIMACRO{\dsum \nolimits_{y}}%
%BeginExpansion
{\displaystyle\sum\nolimits_{y}}
%EndExpansion
p_{y}^{r}\nonumber\\
&  \leq2^{K-t} \label{nk}%
\end{align}
\bigskip where the rational number $r=\frac{2v+1}{2u+1}$, $u$, $v\in N$, and
it is required that $t=rK<2$. The equality of the second inequality in
(\ref{nk}) holds if $p_{y}=2^{-K}$. Similarly, we have $%
%TCIMACRO{\dsum \nolimits_{y^{\prime}}}%
%BeginExpansion
{\displaystyle\sum\nolimits_{y^{\prime}}}
%EndExpansion
(-1)^{W(y^{\prime})}\left\langle \mathbf{I}_{y^{\prime}}^{\prime}\right\rangle
_{LHV}^{r}\leq2^{K-t}$, and hence
\[%
%TCIMACRO{\dsum \nolimits_{y}}%
%BeginExpansion
{\displaystyle\sum\nolimits_{y}}
%EndExpansion
(\left\langle \mathbf{I}_{y}\right\rangle _{LHV}^{r}+(-1)^{W(y)}\left\langle
\mathbf{I}_{y}^{\prime}\right\rangle _{LHV}^{r})\leq2^{K+1-t}.
\]

To find the maximal violation of the nonlinear Bell inequalities, we introduce
the function $f(\theta)=\cos^{t}\theta+\sin^{t}\theta,$ $0<t<2$ and
$0<\theta<\frac{\pi}{2}.$ It is easy to verify that $\frac{d}{d\theta}%
f(\theta)|_{\theta=\frac{\pi}{4}}=0$ and $\frac{d^{2}}{d\theta^{2}}%
f(\theta)|_{\theta=\frac{\pi}{4}}<0$, which leads to
\begin{equation}
\cos^{t}\theta+\sin^{t}\theta\leq\underset{\theta}{\max}(\cos^{t}\theta
+\sin^{t}\theta)=2^{1-\frac{t}{2}}. \label{f}%
\end{equation}
Denote the $K$-bit string $\overline{y}=$ $\overline{y_{1}}\overline{y_{2}%
}\cdots\overline{y_{K}}$, where $\overline{y_{i}}=y_{i}+1$ $\operatorname{mod}%
2$. Let $t=rK$ e have%

\begin{align}
&  \left\langle \mathbf{I}_{y}\right\rangle _{Q}^{r}+\left\langle
\mathbf{I}_{\overline{y}}\right\rangle _{Q}^{r}\nonumber\\
&  =%
%TCIMACRO{\dprod \nolimits_{i=1}^{K}}%
%BeginExpansion
{\displaystyle\prod\nolimits_{i=1}^{K}}
%EndExpansion
(\cos^{\overline{y_{i}}}\theta_{i}\sin^{y_{i}}\theta_{i})^{r}+%
%TCIMACRO{\dprod \nolimits_{i=1}^{K}}%
%BeginExpansion
{\displaystyle\prod\nolimits_{i=1}^{K}}
%EndExpansion
(\cos^{y_{i}}\theta_{i}\sin^{\overline{y_{i}}}\theta_{i})^{r}\nonumber\\
&  =%
%TCIMACRO{\dprod \nolimits_{i=1}^{K}}%
%BeginExpansion
{\displaystyle\prod\nolimits_{i=1}^{K}}
%EndExpansion
(\cos^{rK\overline{y_{i}}}\theta_{i}\sin^{rKy_{i}}\theta_{i})^{\frac{1}{K}}+%
%TCIMACRO{\dprod \nolimits_{i=1}^{K}}%
%BeginExpansion
{\displaystyle\prod\nolimits_{i=1}^{K}}
%EndExpansion
(\cos^{rK\overline{y_{i}}}\theta_{i}\sin^{rK\overline{y_{i}}}\theta
_{i})^{\frac{1}{K}}\nonumber\\
&  =%
%TCIMACRO{\dprod \nolimits_{i=1}^{K}}%
%BeginExpansion
{\displaystyle\prod\nolimits_{i=1}^{K}}
%EndExpansion
\alpha_{i}^{\frac{1}{K}}+%
%TCIMACRO{\dprod \nolimits_{i=1}^{K}}%
%BeginExpansion
{\displaystyle\prod\nolimits_{i=1}^{K}}
%EndExpansion
\beta_{i}^{\frac{1}{K}},\label{y}%
\end{align}
where the condition either $\alpha_{i}=\cos^{t}\theta_{i}$ and $\beta_{i}%
=\sin^{t}\theta_{i}$ or $\alpha_{i}=\sin^{t}\theta_{i}$ and $\beta_{i}=$
$\cos^{t}\theta_{i}$ holds. Next, we exploit the following lemma.

\textit{Lemma 1 }(Mahler inequality) Let $\alpha_{i}$ and $\beta_{i}$ be
nonnegative real numbers; then,%

\begin{equation}
\prod\nolimits_{i=1}^{K}\alpha_{i}^{1/K}+\prod\nolimits_{i=1}^{K}\beta
_{i}^{1/K}\leq\prod\nolimits_{i=1}^{K}(\alpha_{i}+\beta_{i})^{1/K}. \label{g}%
\end{equation}
where the equality holds if $\alpha_{i}=\beta_{i}$ for any $i$. Combining
(\ref{f}), (\ref{y}) and (\ref{g}), we have
\begin{align*}
&  \left\langle \mathbf{I}_{y}\right\rangle _{Q}^{r}+\left\langle
\mathbf{I}_{\overline{y}}\right\rangle _{Q}^{r}\\
&  \leq\left\vert \left\langle \mathbf{I}_{y}\right\rangle _{Q}\right\vert
^{r}+\left\vert \left\langle \mathbf{I}_{\overline{y}}\right\rangle
_{Q}\right\vert ^{r}\\
&  \leq\prod\nolimits_{i=1}^{K}(\cos^{t}\theta_{i}+\sin^{t}\theta_{i})^{1/K}\\
&  \leq2^{1-\frac{t}{2}},
\end{align*}
where all the equalities of these inequalities hold if $\theta_{i}=\frac{\pi
}{4}$ for any $i$. Denote the $K$-bit string $y$ with $y_{1}=0$ by
$y=0y_{2}\cdots y_{K}=0\mathfrak{y}$, and we have
\begin{align}
&  \max%
%TCIMACRO{\dsum \nolimits_{y}}%
%BeginExpansion
{\displaystyle\sum\nolimits_{y}}
%EndExpansion
\left\langle \mathbf{I}_{y}\right\rangle _{Q}^{r}\nonumber\\
&  =\max%
%TCIMACRO{\dsum \nolimits_{\mathfrak{y}}}%
%BeginExpansion
{\displaystyle\sum\nolimits_{\mathfrak{y}}}
%EndExpansion
(\left\vert \left\langle \mathbf{I}_{0\mathfrak{y}}\right\rangle
_{Q}\right\vert ^{r}+\left\vert \left\langle \mathbf{I}_{1\overline
{\overline{\mathfrak{y}}}}\right\rangle _{Q}\right\vert ^{r})\nonumber\\
&  =2^{K-1}(2^{1-\frac{t}{2}})\nonumber\\
&  =2^{K-\frac{t}{2}}, \label{k1}%
\end{align}
and, similarly,
\begin{equation}
\max%
%TCIMACRO{\dsum \nolimits_{y}}%
%BeginExpansion
{\displaystyle\sum\nolimits_{y}}
%EndExpansion
(-1)^{H(y)}\left\langle \mathbf{I}_{y}^{\prime}\right\rangle _{Q}%
^{r}=2^{K-\frac{t}{2}}. \label{k11}%
\end{equation}
We reach the combined non-linear Bell inequality that reads
\begin{equation}
\max%
%TCIMACRO{\dsum \nolimits_{y}}%
%BeginExpansion
{\displaystyle\sum\nolimits_{y}}
%EndExpansion
(\left\langle \mathbf{I}_{y}\right\rangle _{Q}^{r}+(-1)^{H(y)}\left\langle
\mathbf{I}_{y}^{\prime}\right\rangle _{Q}^{r})=2^{K+1-\frac{t}{2}}. \label{k2}%
\end{equation}
As a result, $K$ plays a dominant role in revealing maximum quantum values in
(\ref{k1}-\ref{k2}).

\section{Bell inequalities for ($N$, $K$, $m$) networks}

The star-networks can be extended as the ($N$, $K$, $m$) networks as follows.
There are $N$ independent sources $e_{1}$, $e_{2}$,...,$e_{N}$ and $K+m$
observers/receivers Alice$_{1}$,..., Alice$_{K}$, Bob$_{1}$,..., Bob$_{m}$.
The source $e_{i}$ sends the particles $(2i-1)$ and\ $(2i)$ to Alice$_{i}$ and
Bob$_{i}$, respectively, where $i=1$, $2$, ..., $K$. \ The sources
$e_{i^{\prime}}$ sends two particles $2i^{\prime}-1$ and $2i^{\prime}$ to some
Bob$_{j^{\prime}}$ and Bob$_{j^{\prime\prime}}$, where $K+1\leq i^{\prime}\leq
N$, $j^{\prime}$, $j^{\prime\prime}\in\{1$, $2$,..., $m\}$ and $j^{\prime}%
\neq$ $j^{\prime\prime}$. Therein, Bob$_{j}$ receives $n_{j}$ ($n_{j}\geq2$)
particles with particle indices $j_{1}$, $j_{2}$,..., and $j_{n_{j}}$ from
different $n_{j}$ sources. Without loss of generality, let $j_{1}<$
$j_{2}<\cdots<$ $j_{n_{j}}$, and $j_{1}=2j$ if $1\leq j\leq K$. Denote
Alice$_{i}$ 's input and output by $x_{i}\ $and $a_{x_{i}}^{(i)}$,
respectively, and Bob$_{j}$'s $n_{j}$-bit input string by $y^{(j)}=y_{j_{1}%
}^{(j)}y_{j_{2}}^{(j)}\cdots y_{j_{n_{j}}}^{(j)}$ and one-bit output by
$b_{y^{(j)}}^{(j)}$, where the input bits $y_{j_{1}}^{(j)}$, $y_{j_{2}}^{(j)}%
$, $\cdots$, $y_{j_{n_{j}}}^{(j)}$, $x_{i}$ $\in\{0$, $1\}$ and the output
$b_{y^{(j)}}^{(j)}$, $a_{x_{i}}^{(i)}$ $\in\{1$, $-1\}$. Denote the correlator
$\mathbf{I}_{Y}^{(N,\text{ }K,\text{ }m)}=\frac{1}{2^{K}}%
%TCIMACRO{\dprod \nolimits_{j=1}^{m}}%
%BeginExpansion
{\displaystyle\prod\nolimits_{j=1}^{m}}
%EndExpansion
b_{y^{(j)}}^{(j)}%
%TCIMACRO{\dprod \nolimits_{i=1}^{K}}%
%BeginExpansion
{\displaystyle\prod\nolimits_{i=1}^{K}}
%EndExpansion
(a_{0}^{(i)}+(-1)^{y_{i}}a_{1}^{(i)})$, where $Y$ denotes the $K$-bit string
$y_{1}y_{2}\cdots y_{K},$ $y_{j}=y_{j_{1}}^{(j)}$. In the classical ($N$, $K$,
$m$) networks, the local hidden variable $\lambda=\lambda_{1}\cup\lambda
_{2}\cdots\cup\lambda_{N}$ and shared randomness are distributed throughout
the network. Denote the probability $p_{Y}=%
%TCIMACRO{\dsum \nolimits_{\lambda}}%
%BeginExpansion
{\displaystyle\sum\nolimits_{\lambda}}
%EndExpansion
P(\left\vert \frac{1}{2^{k}}%
%TCIMACRO{\dprod \nolimits_{i=1}^{K}}%
%BeginExpansion
{\displaystyle\prod\nolimits_{i=1}^{K}}
%EndExpansion
(a_{0}^{(i)}+(-1)^{y_{i}}a_{1}^{(i)})%
%TCIMACRO{\dprod \nolimits_{j=1}^{m}}%
%BeginExpansion
{\displaystyle\prod\nolimits_{j=1}^{m}}
%EndExpansion
b_{y^{(j)}}^{(j)}\right\vert =1|\lambda)P(\lambda)$. Consequently, note that
$p_{Y}=\left\langle \left\vert \mathbf{I}_{Y}^{(N,\text{ }K,\text{ }%
m)}\right\vert \right\rangle _{LHV}$ and the first linear Bell-type local
inequality reads%

\begin{align*}
&
%TCIMACRO{\dsum \nolimits_{Y}}%
%BeginExpansion
{\displaystyle\sum\nolimits_{Y}}
%EndExpansion
\left\langle \mathbf{I}_{Y}^{(N,\text{ }K,\text{ }m)}\right\rangle
_{LHV}\text{ }\\
&  \leq%
%TCIMACRO{\dsum \nolimits_{Y}}%
%BeginExpansion
{\displaystyle\sum\nolimits_{Y}}
%EndExpansion
\left\langle \left\vert \mathbf{I}_{Y}^{(N,\text{ }K,\text{ }m)}\right\vert
\right\rangle _{LHV}\text{ }\\
&  =%
%TCIMACRO{\dsum \nolimits_{Y}}%
%BeginExpansion
{\displaystyle\sum\nolimits_{Y}}
%EndExpansion
p_{Y}\text{ }\\
&  =1.
\end{align*}

In the Bell test of the second Bell inequality, denote Alice$_{i}$ 's input
and output by $x_{i}^{\prime}\ $and $a_{x_{i}^{\prime}}^{\prime(i)}$,
respectively, and Bob$_{j}$'s $n_{j}$-bit input string by $y^{\prime
(j)}=y_{j_{1}}^{\prime(j)}y_{j_{2}}^{\prime(j)}\cdots y_{j_{n_{j}}}%
^{\prime(j)}$ and one-bit output by $b_{y^{(j)}}^{\prime(j)}$, where the input
bits $y_{j_{1}}^{\prime(j)}$, $y_{j_{2}}^{\prime(j)}$, $\cdots$, $y_{j_{n_{j}%
}}^{\prime(j)}$, $x_{i}^{\prime}$ $\in\{0$, $1\}$ and the output bits
$b_{y^{\prime(j)}}^{\prime(j)}$, $a_{x_{i}^{\prime}}^{\prime(i)}$ $\in\{1$,
$-1\}$. Denote the correlator $\mathbf{I}_{Y^{\prime}}^{\prime(N,\text{
}K,\text{ }m)}=\frac{1}{2^{K}}%
%TCIMACRO{\dprod \nolimits_{j=1}^{m}}%
%BeginExpansion
{\displaystyle\prod\nolimits_{j=1}^{m}}
%EndExpansion
b_{y^{(j)}}^{(j)}%
%TCIMACRO{\dprod \nolimits_{i=1}^{K}}%
%BeginExpansion
{\displaystyle\prod\nolimits_{i=1}^{K}}
%EndExpansion
(a_{0}^{(i)}+(-1)^{y_{i}}a_{1}^{(i)})$, where $Y^{\prime}$ denotes the $K$-bit
string $y_{1}^{\prime}y_{2}^{\prime}\cdots y_{K}^{\prime},$ $y_{j}^{\prime
}=y_{j_{1}}^{\prime(j)}$. In the classical region, denote the probability
$p_{Y^{\prime}}=%
%TCIMACRO{\dsum \nolimits_{\lambda}}%
%BeginExpansion
{\displaystyle\sum\nolimits_{\lambda}}
%EndExpansion
P(\left\vert \frac{1}{2^{k}}%
%TCIMACRO{\dprod \nolimits_{i=1}^{K}}%
%BeginExpansion
{\displaystyle\prod\nolimits_{i=1}^{K}}
%EndExpansion
(a_{0}^{(i)}+(-1)^{y_{i}^{\prime}}a_{1}^{(i)})%
%TCIMACRO{\dprod \nolimits_{j=1}^{m}}%
%BeginExpansion
{\displaystyle\prod\nolimits_{j=1}^{m}}
%EndExpansion
b_{y^{(j)}}^{(j)}\right\vert =1|\lambda)P(\lambda)$, and the second linear
Bell-type local inequality reads%
\begin{align*}
&
%TCIMACRO{\dsum \nolimits_{Y^{\prime}}}%
%BeginExpansion
{\displaystyle\sum\nolimits_{Y^{\prime}}}
%EndExpansion
(-1)^{W(Y^{\prime})}\left\langle \mathbf{I}_{Y^{\prime}}^{\prime(N,\text{
}K,\text{ }m)}\right\rangle _{LHV}\\
&  \leq%
%TCIMACRO{\dsum \nolimits_{Y}}%
%BeginExpansion
{\displaystyle\sum\nolimits_{Y}}
%EndExpansion
\left\langle \left\vert \mathbf{I}_{1Y}^{(N,\text{ }K,\text{ }m)}\right\vert
\right\rangle _{LHV}\text{)}\\
&  =%
%TCIMACRO{\dsum \nolimits_{Y}}%
%BeginExpansion
{\displaystyle\sum\nolimits_{Y}}
%EndExpansion
p_{Y^{\prime}}\text{ }\\
&  =1.
\end{align*}

In the quantum region, let the observables $A_{x_{i}}^{(i)}=\frac{1}{\sqrt{2}%
}(Z_{i}+(-1)^{x_{i}}X_{i})$ ($A_{x_{i}^{\prime}}^{\prime(i)}=\frac{1}{\sqrt
{2}}(Z_{i}+(-1)^{x_{i}^{\prime}}Y_{i})$ ) and $B_{y^{(j)}}=%
%TCIMACRO{\dprod \nolimits_{k=1}^{n_{j}}}%
%BeginExpansion
{\displaystyle\prod\nolimits_{k=1}^{n_{j}}}
%EndExpansion
Z_{j_{k}}^{\overline{y_{j_{k}}^{(j)}}}X_{j_{k}}^{y_{j_{k}}^{(j)}}$
($B_{y^{(j)}}^{\prime}=%
%TCIMACRO{\dprod \nolimits_{k=1}^{n_{j}}}%
%BeginExpansion
{\displaystyle\prod\nolimits_{k=1}^{n_{j}}}
%EndExpansion
Z_{j_{k}}^{\overline{y_{j_{k}}^{(j)}}}Y_{j_{k}}^{y_{j_{k}}^{(j)}}$) in the
first (second) linear Bell inequality. Hence, the segmented Bell operator for
the first Bell inequality reads%
\begin{equation}
\mathbf{I}_{Y}^{(N,K,m)}=\frac{1}{\sqrt{2^{K}}}%
%TCIMACRO{\dprod \nolimits_{i=1}^{N}}%
%BeginExpansion
{\displaystyle\prod\nolimits_{i=1}^{N}}
%EndExpansion
(Z_{2i-1}Z_{2i})^{\overline{y_{i}}}(X_{2i-1}X_{2i})^{y_{i}},\label{a}%
\end{equation}
and the one for the second Bell inequality reads%

\begin{equation}
\mathbf{I}_{Y^{\prime}}^{\prime(N,K,m)}=\frac{1}{\sqrt{2^{K}}}%
%TCIMACRO{\dprod \nolimits_{i=1}^{N}}%
%BeginExpansion
{\displaystyle\prod\nolimits_{i=1}^{N}}
%EndExpansion
(Z_{2i-1}Z_{2i})^{\overline{y_{i}}}(Y_{2i-1}Y_{2i})^{y_{i}}.\label{aa}%
\end{equation}
And we have
\begin{equation}
\max\{%
%TCIMACRO{\dsum \nolimits_{\alpha}}%
%BeginExpansion
{\displaystyle\sum\nolimits_{\alpha}}
%EndExpansion
\left\langle \mathbf{I}_{\alpha}^{(N,K,m)}\right\rangle _{Q}\}=\sqrt{2^{K}%
},\alpha=Y,Y^{\prime}.\label{10}%
\end{equation}
Finally note that $2N$ operators $\sqrt{2^{K}}\mathbf{I}_{0Y}^{(N,K,m)}$ and
$(-1)^{W(Y^{\prime})}\sqrt{2^{K}}\mathbf{I}_{1Y}^{(N,K,m)}$ with $W(Y)=1$ and
$W(Y^{\prime})=1$ equal to one are linearly independent and each of them can
stabilize the state $%
%TCIMACRO{\dprod \nolimits_{i=1}^{K}}%
%BeginExpansion
{\displaystyle\prod\nolimits_{i=1}^{K}}
%EndExpansion
\left\vert \Phi^{+}\right\rangle _{(2i-1)(2i)}^{\otimes N}$. Hence only $%
%TCIMACRO{\dprod \nolimits_{i=1}^{K}}%
%BeginExpansion
{\displaystyle\prod\nolimits_{i=1}^{K}}
%EndExpansion
\left\vert \Phi^{+}\right\rangle _{(2i-1)(2i)}^{\otimes N}$ can reach the
maximum quantum values (\ref{a}) and (\ref{aa}) simutaneously. We omit the
construction of non-linear Bell inequalities and their maximal violation in
the ($N$, $K$, $m$) networks that are very close to those in the $K$-source
star-networks. In the end, it should be noted that $K$-source star-networks
can be recovered from ($N$, $K$, $m$) networks by setting $N=$ $m=K$ and
Bob$_{1}=\cdots=$Bob$_{K}=$Bob. We omit the proofs of (\ref{a}), (\ref{aa}),
and (\ref{10}) since they are almost the same as those of (\ref{Line1}),
(\ref{Line2}) and (\ref{LinMax}). In addition, one can also construct the
nonlinear Bell inequalities of the ($N$, $K$, $m$) networks, which are very
similar to those in the $K$-source star-networks.

\section{Two-Source scenario revisited}

We revisit the two-source network as follows. Source $e_{1}$ emits a state of
particles 1 and 2, which are sent to Alice and Bob; source $e_{2}$ emits a
state of particles 3, 4, and 5. We consider two different particle
distributions as follows.

\textit{Case (a)} Particles 3 and 4 are sent to Bob and particle 5 is sent to
Charlie. The one-bit inputs for Alice and Charlie are $x$ and $z$,
respectively, and the three-bit string input for Bob is $y_{2}y_{3}y_{4}$. The
output bits for Alice Bob and Charlie are denoted by $a_{x}$, $b_{y_{2}%
y_{3}y_{4}}$, and $c_{z}$, respectively. Consider the correlators
$\mathbf{I}_{y_{2}y_{3}y_{4}}=\frac{1}{4}(a_{0}+(-1)^{y_{2}}a_{1}%
)b_{y_{2}y_{3}y_{4}}(c_{0}+(-1)^{y_{3}+y_{4}+1}c_{1})$, $(y_{2}y_{3}y_{4})\in
s=\{(001),(000),(100),(110)\}$ and $\mathbf{I}_{y_{2}y_{3}y_{4}}^{\prime
}=\frac{1}{4}(a_{0}+(-1)^{y_{2}}a_{1})b_{y_{2}y_{3}y_{4}}^{\prime}%
(c_{0}+(-1)^{y_{3}+y_{4}+1}c_{1})$, $(y_{2}y_{3}y_{4})\in s^{\prime
}=\{(010),(000),(100),(101)\}$ . Similarly, denote the probability
distribution in classical networks%

\begin{align}
&  p_{jk}\nonumber\\
&  =%
%TCIMACRO{\dsum \nolimits_{\lambda}}%
%BeginExpansion
{\displaystyle\sum\nolimits_{\lambda}}
%EndExpansion
P(\left\vert \frac{1}{4}(a_{0}+(-1)^{j}a_{1})(c_{0}+(-1)^{k}c_{1})\right\vert
=1|\lambda)P(\lambda)\nonumber\\
&  =%
%TCIMACRO{\dsum \nolimits_{\lambda}}%
%BeginExpansion
{\displaystyle\sum\nolimits_{\lambda}}
%EndExpansion
P(\left\vert \frac{1}{4}(a_{0}+(-1)^{j}a_{1})b_{y_{2}y_{3}y_{4}}%
(c_{0}+(-1)^{k}c_{1})\right\vert =1|\lambda)P(\lambda)\label{p3}\\
&  =%
%TCIMACRO{\dsum \nolimits_{\lambda}}%
%BeginExpansion
{\displaystyle\sum\nolimits_{\lambda}}
%EndExpansion
P(\left\vert \frac{1}{4}(a_{0}+(-1)^{j}a_{1})b_{y_{2}y_{3}y_{4}}^{\prime
}(c_{0}+(-1)^{k}c_{1})\right\vert =1|\lambda)P(\lambda)
\end{align}
Note that $\left\langle \left\vert \mathbf{I}_{y_{2}y_{3}y_{4}}\right\vert
\right\rangle _{LHV}=\left\langle \left\vert \mathbf{I}_{y_{2}y_{3}y_{4}%
}^{\prime}\right\vert \right\rangle _{LHV}=p_{y_{2}(y_{3}+y_{4}+1)}$, and we
have the first linear Bell inequality%

\begin{align}
&
%TCIMACRO{\dsum \nolimits_{(y_{2}\text{ }y_{3}y_{4})\in s}}%
%BeginExpansion
{\displaystyle\sum\nolimits_{(y_{2}\text{ }y_{3}y_{4})\in s}}
%EndExpansion
\left\langle \mathbf{I}_{y_{2}y_{3}y_{4}}\right\rangle _{LHV}\nonumber\\
&  \leq%
%TCIMACRO{\dsum \nolimits_{(y_{2}\text{ }y_{3}y_{4})\in s}}%
%BeginExpansion
{\displaystyle\sum\nolimits_{(y_{2}\text{ }y_{3}y_{4})\in s}}
%EndExpansion
\left\langle \mathbf{I}_{y_{2}y_{3}y_{4}}\right\rangle _{LHV}\nonumber\\
&  =%
%TCIMACRO{\dsum \nolimits_{j,k=0}^{1}}%
%BeginExpansion
{\displaystyle\sum\nolimits_{j,k=0}^{1}}
%EndExpansion
p_{jk}\nonumber\\
&  \leq1. \label{biGHZ1}%
\end{align}
In the quantum region, the source $e_{1}$ prepares $\left\vert \Phi
^{+}\right\rangle _{12}$, and $e_{2}$ prepares the three-qubit GHZ state
$\left\vert GHZ\right\rangle _{345}$ \ with the density matrix $\left[
GHZ\right]  _{345}=\frac{1}{2^{3}}(I_{3}+X_{3}Z_{4}Z_{5})(I_{3}+Z_{3}%
X_{4}Z_{5})(I_{3}+Z_{3}Z_{4}X_{5})$. Given the input $x$ ($z$), Alice
(Charlie) measures the observable $A_{x}=\frac{1}{\sqrt{2}}(Z_{1}%
+(-1)^{x}X_{1})$ ($C_{z}=\frac{1}{\sqrt{2}}(Z_{5}+(-1)^{z}X_{5})$); given the
input string $y_{1}y_{2}y_{3}$, Bob measures the observable $B_{y_{2}%
y_{3}y_{4}}=%
%TCIMACRO{\dprod \nolimits_{i=2}^{4}}%
%BeginExpansion
{\displaystyle\prod\nolimits_{i=2}^{4}}
%EndExpansion
Z_{i}^{\overline{y_{i}}}X_{i}^{y_{i}}$ or $B_{y_{2}y_{3}y_{4}}^{\prime}=%
%TCIMACRO{\dprod \nolimits_{i=2}^{4}}%
%BeginExpansion
{\displaystyle\prod\nolimits_{i=2}^{4}}
%EndExpansion
Z_{i}^{\overline{y_{i}}}X_{i}^{y_{i}}$ if $(y_{2}y_{3}y_{4})\in s$ or
$(y_{2}y_{3}y_{4})\in s^{\prime}$, respectively. Four independent operators
$2\mathbf{I}_{001}=(Z_{1}Z_{2})(Z_{3}X_{4}Z_{5})$, $2\mathbf{I}_{000}%
=(Z_{1}Z_{2})(Z_{3}Z_{4}X_{5})$, $2\mathbf{I}_{100}=\frac{1}{2}(X_{1}%
X_{2})(Z_{3}Z_{4}X_{5})$, and $2\mathbf{I}_{110}=\frac{1}{2}(X_{1}X_{2}%
)(X_{3}Z_{4}Z_{5})$ stabilize the state $\left\vert \Phi^{+}\right\rangle
_{12}$ $\left\vert GHZ\right\rangle _{345}$. Similarly, there exist five
independent operators that stabilize $\left\vert \Phi^{+}\right\rangle _{12}$
$\left\vert GHZ\right\rangle _{345}$, and hence we propose the second Bell
inequality, which reads%

\begin{align}
&
%TCIMACRO{\dsum \nolimits_{(y_{2}\text{ }y_{3}y_{4})\in s^{\prime}}}%
%BeginExpansion
{\displaystyle\sum\nolimits_{(y_{2}\text{ }y_{3}y_{4})\in s^{\prime}}}
%EndExpansion
\left\langle \mathbf{I}_{y_{2}y_{3}y_{4}}^{\prime}\right\rangle _{LHV}%
\nonumber\\
&  \leq%
%TCIMACRO{\dsum \nolimits_{(y_{2}\text{ }y_{3}y_{4})\in s^{\prime}}}%
%BeginExpansion
{\displaystyle\sum\nolimits_{(y_{2}\text{ }y_{3}y_{4})\in s^{\prime}}}
%EndExpansion
\left\langle \left\vert \mathbf{I}_{y_{2}y_{3}y_{4}}^{\prime}\right\vert
\right\rangle _{LHV}\nonumber\\
&  =%
%TCIMACRO{\dsum \nolimits_{j,k=0}^{1}}%
%BeginExpansion
{\displaystyle\sum\nolimits_{j,k=0}^{1}}
%EndExpansion
p_{jk}\nonumber\\
&  \leq1. \label{biGHZ2}%
\end{align}
It is to verify that these four independent operators $2\mathbf{I}%
_{010}^{\prime}=(Z_{1}Z_{2})(X_{3}Z_{4}Z_{5})$, $2\mathbf{I}_{000}^{\prime
}=2\mathbf{I}_{000}$, $2\mathbf{I}_{100}^{\prime}=2\mathbf{I}_{000}$, and
$2\mathbf{I}_{101}^{\prime}=(X_{1}X_{2})(Z_{3}X_{4}Z_{5})$ also stabilize the
state $\left\vert \Phi^{+}\right\rangle _{12}$ $\left\vert GHZ\right\rangle
_{345}$. Combining (\ref{biGHZ1}) and (\ref{biGHZ2}), we derive the linear
Bell inequality that reads%
\[%
%TCIMACRO{\dsum \nolimits_{(y_{2}\text{ }y_{3}y_{4})\in s}}%
%BeginExpansion
{\displaystyle\sum\nolimits_{(y_{2}\text{ }y_{3}y_{4})\in s}}
%EndExpansion
\left\langle \mathbf{I}_{y_{2}y_{3}y_{4}}\right\rangle +%
%TCIMACRO{\dsum \nolimits_{(y_{2}\text{ }y_{3}y_{4})\in s^{\prime}}}%
%BeginExpansion
{\displaystyle\sum\nolimits_{(y_{2}\text{ }y_{3}y_{4})\in s^{\prime}}}
%EndExpansion
\left\langle \mathbf{I}_{y_{2}y_{3}y_{4}}^{\prime}\right\rangle \overset
{LHV}{\leq}2.
\]
On the other hand, we have
\[
\max\{%
%TCIMACRO{\dsum \nolimits_{(y_{2}\text{ }y_{3}y_{4})\in s}}%
%BeginExpansion
{\displaystyle\sum\nolimits_{(y_{2}\text{ }y_{3}y_{4})\in s}}
%EndExpansion
\left\langle \mathbf{I}_{y_{2}y_{3}y_{4}}\right\rangle _{Q}+%
%TCIMACRO{\dsum \nolimits_{(y_{2}\text{ }y_{3}y_{4})\in s^{\prime}}}%
%BeginExpansion
{\displaystyle\sum\nolimits_{(y_{2}\text{ }y_{3}y_{4})\in s^{\prime}}}
%EndExpansion
\left\langle \mathbf{I}_{y_{2}y_{3}y_{4}}^{\prime}\right\rangle _{Q}\}=4,
\]
which can be achieved using the state $\left\vert \Phi^{+}\right\rangle _{12}$
$\left\vert GHZ\right\rangle _{345}$.

\textit{Case (b)} In this case, particles 3, 4, and 5 are sent to Bob,
Charlie$_{1}$, and Charlie$_{2}$, respectively. Denote the one-bit inputs for
Alice, Charlie$_{1}$, and Charlie$_{2}$ by $x$, $z_{1}$, and $z_{2}$; the
one-bit outputs by $a_{x}^{(1)}$, $c_{z_{1}}^{(4)}$ and $c_{z_{2}}^{(5)}$,
respectively. Denote the two-bit string input and one-bit output for Bob by
$y_{2}y_{3}$ and $b_{y_{2}y_{3}}$, respectively. In classical networks, we
introduce the expectation values of the correlators $\left\langle
\mathbf{I}_{y_{2}y_{3}}\right\rangle =\frac{1}{8}\left\langle (a_{0}%
^{(1)}+(-1)^{y_{2}}a_{1}^{(1)})b_{y_{2}y_{3}}(c_{0}^{(4)}+(-1)^{y_{2}}%
c_{1}^{(4)})(c_{0}^{(5)}+(-1)^{\overline{y_{2}+y_{3}}}c_{1}^{(5)}%
)\right\rangle $ and $\left\langle \mathbf{I}_{y_{2}y_{3}}^{\prime
}\right\rangle =\frac{1}{8}\left\langle (a_{0}^{(1)}+(-1)^{y_{2}}a_{1}%
^{(1)})b_{y_{2}y_{3}}(c_{0}^{(4)}+(-1)\overline{^{y_{2}}}c_{1}^{(4)}%
)(c_{0}^{(5)}+(-1)^{y_{2}+y_{3}}c_{1}^{(5)})\right\rangle $. Similarly, denote
the probability distribution in classical networks%

\begin{align}
&  p_{jkl}\nonumber\\
&  =%
%TCIMACRO{\dsum \nolimits_{\lambda}}%
%BeginExpansion
{\displaystyle\sum\nolimits_{\lambda}}
%EndExpansion
P(\left\vert \frac{1}{8}(a_{0}^{(1)}+(-1)^{j}a_{1}^{(1)})(c_{0}^{(4)}%
+(-1)^{k}c_{1}^{(4)})(c_{0}^{(5)}+(-1)^{l}c_{1}^{(5)})\right\vert
=1|\lambda)P(\lambda)\nonumber\\
&  =%
%TCIMACRO{\dsum \nolimits_{\lambda}}%
%BeginExpansion
{\displaystyle\sum\nolimits_{\lambda}}
%EndExpansion
P(\left\vert \frac{1}{8}(a_{0}^{(1)}+(-1)^{j}a_{1}^{(1)})(c_{0}^{(4)}%
+(-1)^{k}c_{1}^{(4)})(c_{0}^{(5)}+(-1)^{l}c_{1}^{(5)})b_{y_{2}y_{3}%
}\right\vert =1|\lambda)P(\lambda), \label{p4}%
\end{align}
and we have $\left\langle \left\vert \mathbf{I}_{y_{2}y_{3}}\right\vert
\right\rangle _{LHV}=p_{y_{2}y_{2}\overline{y_{2}+y_{3}}}$ and $\left\langle
\left\vert \mathbf{I}_{y_{2}y_{3}}^{\prime}\right\vert \right\rangle
_{LHV}=p_{^{y_{2}\overline{y_{2}}y_{2}+y_{3}}}$. The Bell-type inequality
reads
\begin{align}
&
%TCIMACRO{\dsum \nolimits_{y_{2},\text{ }y_{3}}}%
%BeginExpansion
{\displaystyle\sum\nolimits_{y_{2},\text{ }y_{3}}}
%EndExpansion
((-1)^{y_{2}y_{3}}\left\langle \mathbf{I}_{y_{2}y_{3}}\right\rangle
_{LHV}+(-1)^{\overline{y_{2}}y_{3}}\left\langle \mathbf{I}_{y_{2}y_{3}%
}^{\prime}\right\rangle _{LHV})\nonumber\\
&  \leq%
%TCIMACRO{\dsum \nolimits_{j,k,l}}%
%BeginExpansion
{\displaystyle\sum\nolimits_{j,k,l}}
%EndExpansion
p_{jkl}\nonumber\\
&  =1. \label{ll}%
\end{align}

In the quantum region, given the input $x$ ($z_{1},$ $z_{2}$), Alice
(Charlie$_{1}$, Charlie$_{2}$) measures the observable $A_{x_{1}}^{(1)}%
=\frac{1}{\sqrt{2}}(Z_{1}+(-1)^{x_{1}}X_{1})$ ($C_{z_{1}}^{(4)}=\frac{1}%
{\sqrt{2}}(Z_{4}+(-1)^{z_{1}}X_{4})$, $C_{z_{2}}^{(5)}=\frac{1}{\sqrt{2}%
}(Z_{5}+(-1)^{z_{2}}X_{5})$); given the input string $y_{1}y_{2}$, Bob
measures the observable $B_{y_{2}y_{3}}=%
%TCIMACRO{\dprod \nolimits_{i=2}^{3}}%
%BeginExpansion
{\displaystyle\prod\nolimits_{i=2}^{3}}
%EndExpansion
Z_{i}^{\overline{y_{i}}}X_{i}^{y_{i}}.$ Denote the correlator operators
$\mathbf{I}_{y_{2}y_{3}}=\frac{1}{8}(A_{0}^{(1)}+(-1)^{y_{2}}A_{1}%
^{(1)})B_{y_{2}y_{3}}(C_{0}^{(4)}+(-1)^{y_{2}}C_{1}^{(4)})(C_{0}%
^{(5)}+(-1)^{\overline{y_{2}+y_{3}}}C_{1}^{(5)})$ and $\mathbf{I}_{y_{2}y_{3}%
}^{\prime}=\frac{1}{8}(A_{0}^{(1)}+(-1)^{y_{2}}A_{1}^{(1)})B_{y_{2}y_{3}%
}(C_{0}^{(4)}+(-1)^{\overline{y_{2}}}C_{1}^{(4)})(C_{0}^{(5)}+(-1)^{y_{2}%
+y_{3}}C_{1}^{(5)})$. These eight opertors $2\sqrt{2}\mathbf{I}_{y_{1}y_{2}}$
and $2\sqrt{2}\mathbf{I}_{y_{1}y_{2}}^{\prime}$ $\in\{Z_{1}Z_{2},X_{1}%
X_{2}\}\otimes\{Z_{3}Z_{4}X_{5},Z_{3}X_{4}Z_{5},X_{3}Z_{4}Z_{5}-X_{3}%
X_{4}X_{5}\}$ stabilze the state $\left\vert \Phi^{+}\right\rangle _{12}$
$\left\vert GHZ\right\rangle _{345}$. As a result,%

\[
\max%
%TCIMACRO{\dsum \nolimits_{y_{1},\text{ }y_{2}}}%
%BeginExpansion
{\displaystyle\sum\nolimits_{y_{1},\text{ }y_{2}}}
%EndExpansion
((-1)^{y_{2}y_{3}}\left\langle \mathbf{I}_{y_{2}y_{3}}\right\rangle
_{Q}+(-1)^{\overline{y_{2}}y_{3}}\left\langle \mathbf{I}_{y_{2}y_{3}}^{\prime
}\right\rangle _{Q})=2\sqrt{2},
\]
which can be achieved using the state $\left\vert \Phi^{+}\right\rangle _{12}$
$\left\vert GHZ\right\rangle _{345}$.

In conclusion, we investigate Bell nonlocality with variant particle
distributions in the networks. As for genuine Bell locality, local hidden
variables and randomness can be perfectly cloned and spread throughout the
networks. In quantum networks, each quantum source emits the two-qubit Bell
states. Instead of brutal searching for the maximal classical correlation
strengths, we propose the linear and nonlinear Bell inequalities based on the
probability normalization of non-contextual and predetermined probability
distributions. To find the maximal violations of these proposed Bell
inequalities, the Pauli string operators as the stabilizing operators are
quite useful in assigning the local observables and designing the segmented
Bell operators. As a result, these proposed Bell inequalities can be maximally
violated by Bell states distributed in the quantum networks.

In the upcoming study, we will explore Bell nonlocality with stabilizer states
such as graph states \cite{10,11} and codewords of stabilizer-based quantum
error correction distributed in quantum networks \cite{9}. The local
observables therein can be made up of \textquotedblleft
cut-graft-mixing\textquotedblright\ stabilizing operators and logical
operators \cite{10}. In addition, we seek Bell inequalities tailored for the
non-maximal entangled states distributed in the networks and the trade-off of
randomness and nonlocality therein \cite{random,last}. In addition, we will
further study the steering effect\ and the self-test in the networks
\cite{steer,self}.

\section{Acknowledgment}

The work was supported by Grant No. MOST 110-2112-M-033 -006.

\end{document}